\documentclass[%
 aip,
 jmp,%
 amsmath,amssymb,
 reprint,
]{revtex4-2}

\usepackage{graphicx}% Include figure files
\usepackage{dcolumn}% Align table columns on decimal point
\usepackage{bm}% bold math
\usepackage{braket}
\usepackage{ulem}
\usepackage{siunitx}
\usepackage{hyperref}
\hypersetup{hidelinks}
\hypersetup{colorlinks = true, linkcolor=blue, urlcolor=blue,citecolor=blue}

\begin{document}

\preprint{AIP/123-QED}

\title{Quantum Simulations with Bilayer 2D Bose Gases in Multiple-RF-dressed Potentials}

\author{Abel Beregi}
\affiliation{Clarendon Laboratory, University of Oxford, Oxford OX1 3PU, United Kingdom}
\author{Christopher Foot}%
\affiliation{Clarendon Laboratory, University of Oxford, Oxford OX1 3PU, United Kingdom}
\author{Shinichi Sunami}%
\email{shinichi.sunami@physics.ox.ac.uk}
\affiliation{Clarendon Laboratory, University of Oxford, Oxford OX1 3PU, United Kingdom}

\date{\today}

\begin{abstract}
Multiple-RF (MRF) dressing allows trapping of ultracold atoms in novel spatial geometries, such as highly controllable bilayer structures for 2D ultracold gases, providing unique opportunities for the investigation of 2D quantum systems both in and out of equilibrium. 
Here, we give an overview of the recent developments of MRF-dressed atom experiments, illustrated by the detailed studies of universal relaxation dynamics across the Berezinskii-Kosterlitz-Thouless critical point enabled by coherent splitting quench protocols and detection of correlations via spatially selective matter-wave interferometry.

\end{abstract}

\maketitle

\section{\label{sec:intro} Introduction}

Ultracold quantum gases have emerged as clean and highly controllable experimental platforms suitable for use as quantum simulators to study the equilibrium and dynamical properties of quantum many-body systems, which may be beyond the reach of existing numerical simulation techniques \cite{Gross2017,Schafer2020}. 
The high controllability of ultracold atomic systems relies largely on the flexibility of the spatial trapping geometries that are made possible by ingenious application and control of electromagnetic fields spanning from DC to optical frequencies. 
The use of optical dipole forces \cite{Grimm1999} is a common way to create a wide range of geometries including lattice\cite{Deutsch1998}, box\cite{Gaunt2013a}, and disordered potentials \cite{Lye2005}.

Magnetic trapping offers an alternative regime of controllability for confinement and manipulation of atoms thanks to the differing spatial constraints.
In particular, trapping based on the adiabatic dressed states of atoms with strong radio-frequency (RF) radiation, the RF-dressed potential technique \cite{Zobay2001}, realizes unique topologies including chip-based elongated double-well traps \cite{Schumm2005, Jo2007, vanEs2008, Baumgartner2010, Kim2016}, bubble traps \cite{Carollo2022}, toroidal traps \cite{Lesanovsky2006, Lesanovsky2007, Heathcote2008, Kim2016, Navez2016} and two-dimensional (2D) potentials \cite{Colombe2004,Sunami2022}, as well as enhancing the capability of other types of atom traps by RF-dressed optical lattice \cite{Lundblad2008} and microwave-field dressing \cite{Bohi2009, Ammar2015}. 

Multiple-RF (MRF) dressing \cite{Harte2018} significantly enhances the capability of RF-dressed potentials, allowing the generation of programmable potentials such as multi-well structures with well-selective trap control and probing \cite{Luksch2019, Barker2020}, species-selective trapping and manipulation \cite{Bentine2017, Bentine2020, Barker2020jphysb}, including the rapid and reproducible manipulation of the potential \cite{Barker2020, Sunami2023}.
MRF-dressing of atoms in a quadrupole magnetic field can produce a bilayer system which is a very powerful experimental platform for investigating the properties of 2D quantum systems via matter-wave interferometry readout of the local phase observable, including a rich variety of non-equilibrium behavior such as the dynamics of Berezinskii-Kosterlitz-Thouless superfluid \cite{Berezinskii1972, Kosterlitz1973, Mathey2007, Hadzibabic2006, Sunami2022, Sunami2023}.
The MRF-dressed bilayer structure further presents opportunity to probe systems with inter-layer coupling, by precisely tuning the inter-layer distance by the choice of RF frequencies used to trap the atoms. 
The coupled bilayer system provides a flexible platform to study layered superconductors, such as counterflow \cite{Homann2024, Anzi2009} and light-induced \cite{Zhu2021,Okamoto2016} superfluidity, vortex dynamics of coupled field \cite{Eto2018, Kobayashi2019}, quantum field theories \cite{Viermann2022, Zache2020}, and non-equilibrium dynamics of 2D quantum systems following a coupling quench \cite{Mathey2007, Mathey2017}.

In this perspective, we review the theoretical background, experimental implementation and applications of MRF dressing, focusing on the generation, manipulation and detection of bilayer 2D quantum gases and provide prospect of these techniques to probe a wider class of quantum many-body systems.
Firstly, in Sec.~\ref{sec:mrf}, we give a brief introduction to MRF-dressed potentials and their experimental implementation with a specific static magnetic field configuration (quadrupole field), which generates bilayer 2D traps for ultracold atoms.
Next, in Sec.~\ref{sec:bkt}, we discuss the matter-wave interferometry detection of bilayer 2D quantum gases, enabled by the versatile MRF-dressed double-well potential, and illustrate its application to the direct statistical analysis of phase correlations.
Sec.~\ref{sec:quench} describes a quench protocol for 2D quantum gases realized by MRF dressing, used to probe the universal relaxation dynamics following a quench across the BKT critical point.
Finally, in Sec.~\ref{sec:outlook}, we provide an outlook for the MRF-dressing technique beyond the bilayer structure, outlining the application to probe layered structures in unique topologies and multi-species systems with individually controllable trap geometries.

\section{\label{sec:mrf} MRF-dressed potentials}
The fundamental concept underlying RF-dressed potentials is magnetic resonance between Zeeman sublevels \cite{Zobay2001, Perrin2017}.
For a weak external static magnetic field\footnote{This is a valid assumption for typical RF-dressed setups as the RF-frequencies are in the MHz range while the hyperfine splitting is of the GHz order of magnitude. For higher radio-frequencies, the influence of the non-linear Zeeman effect on dressed potentials is discussed in \cite{Sinuco-Leon2012}}, the eigenenergies are given by $E_{m_F} = m_F g_F \mu_B |\bm{B}|$. 
With an additional AC magnetic field $\bm{B}_{\text{AC}}=B_{AC} \sin (\omega t + \phi) \hat{\bm{\epsilon}}$, magnetic resonance occurs at $|\bm{B}| = \hbar \omega / |g_F| \mu_B$ with a strength that depends on the orientation of the polarization unit vector $\hat{\bm{\epsilon}}$ relative to the local static field.
Assuming a constant magnetic field gradient along $z$, $B_{DC} = - z b_z$, such as in quadrupole field configuration along the $z$ axis, we seek the eigenenergies of the semi-classical Hamiltonian
\begin{equation}
    \hat{\mathcal{H}} =  \frac{g_F \mu_B}{\hbar} \left( B_{DC} \hat{F}_z + \bm{B}_{AC} \cdot \hat{\bm{F}} \right) , 
    \label{eq:rf_dressed_hamiltonian}
\end{equation}
where $\hat{\bm{F}}$ is the operator for the total atomic angular momentum. 
In the rotating wave approximation, the eigenenergies are given by \cite{Perrin2017}
\begin{equation}
    E = \widetilde{m}_F \hbar \sqrt{\delta^2+\Omega^2},
    \label{eq:dressed_energies}
\end{equation}
where $\widetilde{m}_F$ is a new quantum number which takes the same values as $m_F$ for a given species, $\delta = \omega - |g_F| \mu_B B_{DC}/\hbar$ is the angular detuning from resonance, and $\Omega$ is the Rabi frequency. 
The eigenenergy of the state $\ket{\widetilde{m}_F=1}$ has a local minimum on resonance and this serves as the trapping state. 
For typical experimental parameters in 2D Bose gas experiments, the atomic motion satisfies the adiabatic condition to follow the same dressed eigenstate \cite{Perrin2017, Burrows2017}, allowing long lifetime in the trap.
Since the trap minimum is located close to an isomagnetic surface of the static magnetic field, this results in a bubble-shaped trap of aspect ratio 2 for the quadrupole field. 
Such potentials are highly anisotropic: 
perpendicular to the bubble, the strong confinement is provided by the energy cost of detuning tangentially to the surface of the bubble, and a weak restoring force comes from the effect of gravity to confine atoms near the bottom \footnote{The spatial variations of the Rabi-frequency also contribute to the total force, but in most scenarios, their effect is small.}, thus confining quantum gases in the quasi-2D regime.

To understand MRF-dressing, we extend the theory of adiabatic potentials to AC magnetic fields of multiple tones, $\bm{B}_{\text{AC}}=\sum B_{AC}^{(i)} \sin (\omega_i t + \phi_i) \hat{\bm{\epsilon}}_i$. 
The Hamiltonian in this case is of the same form as in Eq. (\ref{eq:rf_dressed_hamiltonian}), but with the multi-tone RF-field. 
Diagonalization of the MRF-dressed Hamiltonian is possible by employing non-degenerate bases \cite{Luksch2019}, however, the computational cost is significant even for three RF components. An alternative method of obtaining MRF-dressed eigenenergies is the application of a semi-classical Floquet theory, provided that the Hamiltonian is time-periodic\footnote{This is the case for the experimentally relevant signals derived from a common fundamental.}. 
In this case, the Schr\"{o}dinger equation is integrated numerically over one period and the eigenvalues can be extracted from the complex phase accumulated over this evolution \cite{Shirley1965}.
The open-source Matlab software library CalcAP \cite{crfap, BentineThesis} provides a comprehensive interface to calculate MRF and time-averaged adiabatic potentials from the RF and DC magnetic field configurations including, on 1D, 2D and 3D grid mapping, which is extensively unit-tested and confirmed with several experimental results \cite{Harte2018, Barker2020}. 

\begin{figure}[t]
	\includegraphics[width=0.86\textwidth]{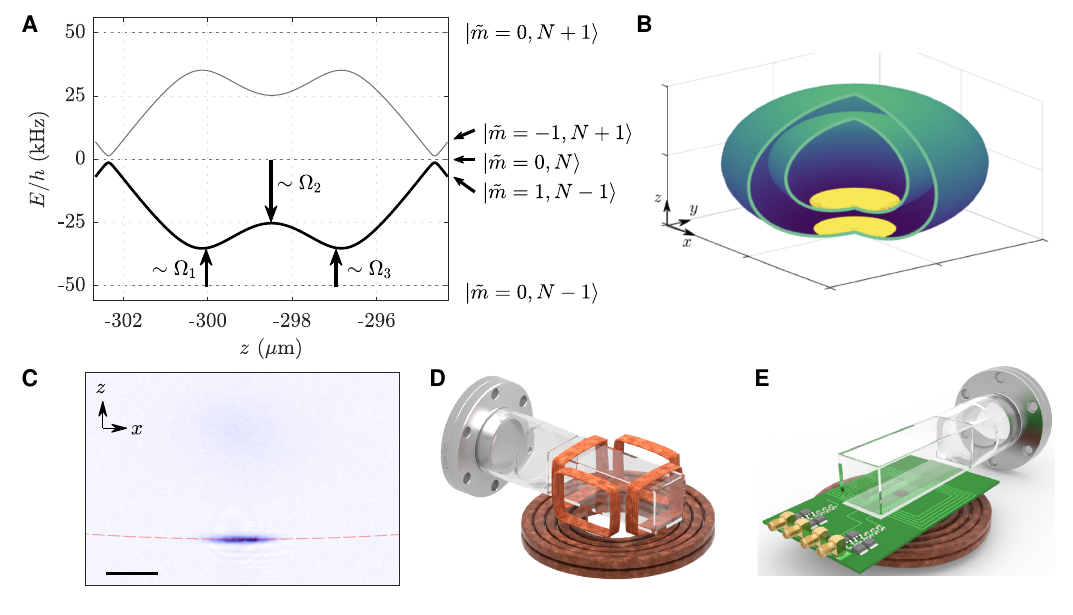}
	\caption{ MRF-dressed potential and experimental implementation for the realization of bilayer 2D systems.
		(\textbf{A}) Dressed eigenenergies of an MRF-dressed potential obtained from numerical simulation \cite{Harte2018,crfap} for typical experimental parameters: $\omega_i/2\pi = \{ 7.1, 7.15, 7.2 \}$ MHz with $B_{AC}^{(i)} = \{ 49, 93, 49 \}$ mG, corresponding to Rabi-frequencies of $\Omega_{i}/2\pi = \{34, 65, 34\}$ kHz, and a quadrupole gradient of 173.5 G/cm.
		With these parameters, the double-well potential is formed for the $\ket{\tilde{m}=1, N-1}$ dressed state (bold black line) which has a well separation of $ \SI{2.8}{ \micro \meter}$ at $z \approx \SI{-300}{\micro \meter}$ below the quadrupole node. 
		Additional, higher-order atom-photon interactions result in weak avoided crossings at $z \approx \SI{-302}{\micro \meter}$ and $\SI{-294}{\micro \meter}$. 
		%Dotted lines indicate the magnetically insensitive states and the single-well potential for the $\ket{\tilde{m}=-1, N+1}$ state is plotted with a solid line. 
		(\textbf{B}) Illustration of MRF-dressed potential with quadrupole static field, resulting in double spheroidal surfaces on which atoms are confined (the difference in size of the spheroids is exaggerated).
		(\textbf{C}) Absorption image of atoms trapped in a single-RF dressed potential \cite{SunamiThesis, Barker2020}.
		The typical diameter of the 2D system is tens of micrometers, while the resonant spheroid (red dashed line) has a major axis of $\sim \SI{600}{\micro \meter}$, thus the atoms are confined along a flat geometry.
		Black bar denotes $\SI{30}{\micro \meter}$.
		(\textbf{D}, \textbf{E}) Examples of coil array geometries for MRF-dressed quadrupole trap \cite{Harte2018,BeregiThesis}. 
		The DC quadrupole field is generated by high current coils below and above (not shown) the vacuum cell, while the RF fields are generated, for example, by pairs of coils \cite{Luksch2019} (connected in a quasi-Helmholtz configuration for each pair) (\textbf{D}) or flat array on a printed circuit board \cite{BeregiThesis} (\textbf{E}).
	}
	\label{fig:trap}
\end{figure}

The manipulation of the MRF-dressed potential is achieved by the control of RF signal amplitudes $B_{AC}^{(i)}$, as illustrated in Fig.~\ref{fig:trap}A, where we plot states with different Fock number $N$ for the RF field \cite{Harte2018} for clarity.
Amplitudes of the RF components $\omega_1$ and $\omega_3$ control the imbalance of the two wells, while $\Omega_2$ determines the barrier height and hence how the overall trap shape varies from single to double-well potentials.

\subsection{Experimental implementation with a quadrupole DC magnetic field}\label{sec:impl}

An MRF-dressed trap has been demonstrated in a static quadrupole field, resulting in layered 2D quantum gases\cite{Harte2018,Bentine2017,Sunami2022,Sunami2023,BeregiThesis}.
The trap for these experiments was formed using the coils shown in Figure \ref{fig:trap}D: high-current anti-Helmholtz coils generating a DC quadrupole field, and quasi-Helmholtz pairs for the independently controllable AC fields in $x$ and $y$ directions.
In these works, AC coils were driven by multiple tones of RF signals generated by DDS chips (with a common clock signal), ensuring stable relative phases between frequency components
\footnote{Dressing with multiple radio-frequencies results in complex higher-order atom-photon interactions which cause atom loss, therefore particular attention is required for the design of the RF signal chain. Firstly, it is important to use commensurate RF frequencies as non-linear behavior of components in the RF signal chain results in intermodulation products. Practically, resonant matching of the RF coils to the source impedance provides stable RF amplitudes and filtering of unwanted frequency components.}.

As RF-dressed traps have an inherently highly anisotropic potential, efficient loading from a spherical distribution of laser-cooled atoms in a Magneto-Optical Trap (MOT) requires careful sequence design to ensure mode-matching.
The first implementation of an RF-dressed trap relied on loading ultracold atoms from a type of Ioffe-Pritchard trap (QUIC trap) \cite{Colombe2004}, by sweeping the frequency of the dressing RF field from below the Larmor-frequency at the center of the trap to a value above, thus smoothly deforming the potential. 
Another scheme \cite{BentineThesis,Harte2018}, in which the radio-frequency remains fixed, involves the successive transfer of atoms from a quadrupole trap, time-orbiting potential \cite{Petrich1995}, then into time-averaged adiabatic trap (TAAP) \cite{Gildemeister2010, Lesanovsky2007} and finally into the RF-dressed trap by reducing the strength of the rotating bias field. 
A fixed radio-frequency allows the usage of resonantly matched RF-antennae to ensure strong AC magnetic field generation and filtering of undesired frequency components that induce atom losses \cite{Luksch2019}.
In a recent technical development \cite{BeregiThesis}, aided by automated Bayesian optimization \cite{Barker2020ml}, loading of atoms into an RF-dressed potential by a near-instantaneous projection into a trapped RF-dressed manifold from the bare Zeeman states of atoms in optical molasses was demonstrated \cite{BeregiThesis}; 
this efficient loading of thermal clouds and subsequent evaporative cooling result in stable production of Bose condensates at the bottom of the RF-dressed potential without the necessity of dynamically loading the condensates in to the dressed potential, which may excite collective modes if the process is not fully adiabatic.

From the single-RF dressed potential, ultracold atoms can be adiabatically loaded into an MRF-dressed double well through a series of transformations \cite{SunamiThesis, Barker2020, Sunami2023}.
The first stage is projection from a single-RF trap to the three-RF MRF trap with amplitude combination chosen to give a single-well potential which is mode-matched to the original trap; 
this is achieved by maintaining the middle frequency component $\omega_2$ with a large amplitude similar to that of the original single-RF trap, while the other two RF components $\omega_{1}$ and $\omega_{3}$ are switched on at a small amplitude. 
Secondly, the amplitudes of all RF components $\Omega_i$ are ramped to realize a flat-bottom potential. The final stage is adiabatic splitting into a double-well, by reducing $\Omega_2$ while ramping $\Omega_1$ and $\Omega_3$ carefully in a way that ensures a balanced double well throughout the process  \footnote{%Something to generalise the exp methods
	While such a splitting protocol requires precise adjustment of individual field amplitudes at various stages, the integration of high-precision calibrations through RF-spectroscopy \cite{Easwaran2010} and the capability to calculate the potential from recorded waveforms make this scheme feasible.}.
Typically, slow ramps of duration up to a second are used to prepare equilibrated sample of bilayer 2D Bose gases, while this process can be significantly accelerated to achieve rapid \textit{coherent splitting} quench of the system \cite{Barker2020}, as described in Sec.~\ref{sec:quench}.

\begin{figure}[t]
\includegraphics[width=0.7\textwidth, trim={0 0 4.27cm 0},clip]{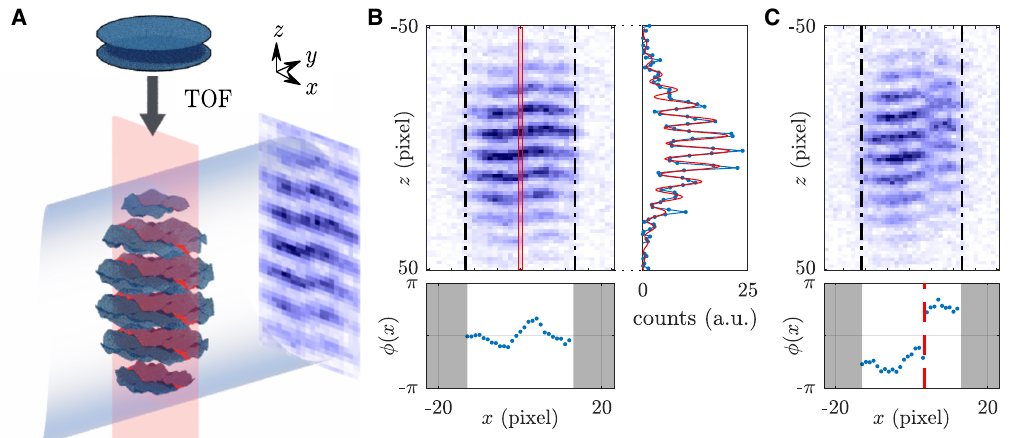}
\caption{\label{fig:fringe} Probing phase fluctuations of bilayer 2D Bose gases using matter-wave interferometry, data taken from Ref.\ \cite{Sunami2022, Sunami2023}.
(\textbf{A}) Illustration of the experimental procedure. 
We begin with quasi-2D bilayer Bose gases in an MRF double-well trap (blue discs, top). 
After time-of-flight (TOF) expansion, the clouds spatially overlap to produce interference fringes with fluctuating phases (blue wavy planes). 
The red shading denotes the thin light sheet that optically pumps the atoms to the $F=2$ hyperfine level, which we image using light resonant with the cycling transition (depicted as a blue beam propagating along the $y$ direction).	
(\textbf{B}) The obtained images are analyzed column by column (illustrated by red box on the image, with corresponding one-dimensional data on the right together with the fit), obtaining the spatial phase profile plotted below.
}
\end{figure}

\begin{figure}[t]
	\includegraphics[width=0.88\textwidth]{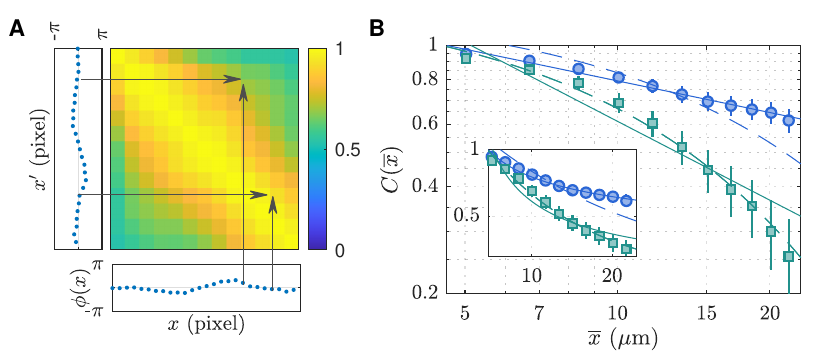}
	\caption{\label{fig:corrfunc} Phase correlation functions of 2D Bose gases, data from Ref.\ \cite{BeregiThesis}.
		(\textbf{A}) An example of the correlation function $C^r(x,x')$ computed from 50 realizations of $\phi(x)$, see Fig.~\ref{fig:fringe} for the extraction of $\phi(x)$ from the interference images. 
		Diagonal elements $C^r(x,x)$ are always unity by definition. 
		(\textbf{B}) From $C^r(x,x')$, the correlation function $C(\overline{x})$ at distance $\overline{x}$ is calculated by taking the mean of $\overline{x}$-th diagonal elements of $C^r(x,x')$. 
		Here, we show the correlation function $C(\overline{x})$ from two separate datasets taken at two temperatures below (blue circle) and above (green square) the BKT crossover, with near-homogeneous 2D Bose gases realized by an additional optical ring potential \cite{BeregiThesis}. 
		The solid and dashed lines are the fits with a power-law and exponential models respectively.
	}
\end{figure}

\section{\label{sec:bkt} Matter-wave interferometry of 2D Quantum gases}

The spatially resolved interferometry of low-dimensional quantum gases provides a powerful tool to understand the fundamental properties of correlated many-body states, such as 1D quasicondensates in and out of equilibrium \cite{Hofferberth2007, Langen2013, Schweigler2017, Zache2020}.
Recent realization of a bilayer trap enabled by MRF-dressed potentials \cite{Barker2020, Sunami2022} demonstrated an advancement from the early pioneering work by Hadzibabic \textit{et al.} \cite{Hadzibabic2006, Kruger2007} using matter-wave interferometry to identify the fluctuations present in 2D Bose gases including non-equilibrium systems.
The major advances in recent work are the possibility to perform precise dynamical trap manipulation with MRF-dressed potentials, as well as the realization of spatially selective detection of interfering clouds allowing the extraction of locally resolved phase fluctuations, which leads to direct analysis of two-point and higher-order correlation functions \cite{Rath2010, Sunami2022, Zache2020}, crucial observable for the comprehensive understanding of novel many-body quantum systems.

The experimental procedure is illustrated in Fig.~\ref{fig:fringe}A.
The trap is turned off abruptly by fast RF switches and matter-wave interference fringes form after a time-of-flight expansion.
A slice of the density distribution is imaged by a spatially selective repumping scheme, as illustrated in the figure.
Absorption imaging of the repumped atoms along the $y$ axis yields images as shown in Fig.~\ref{fig:fringe}B which display spatially fluctuating phases.
%supports both thermal and superfluid phases of the BKT transition \cite{Clade2009,Posazhennikova2006,Prokofev2002,Druten1997}.
%
The local fluctuations of the interference fringes contain information about the relative phases of the \textit{in situ} clouds since they do not expand significantly in the radial direction during time-of-flight.
At each location $x$, the interference pattern can be fitted
%with the function \cite{Pethick2008}
%%
%\begin{equation}\label{eq:fringefit}
%\rho_x(z) = \rho_0 \exp\left(-z^2/2\sigma^2\right) \left[ 1 + c_0 \cos(kz+\phi(x)) \right],
%\end{equation}
%
%where $\rho_0,\sigma, c_0,k,\phi(x)$ are fit parameters. The 
to extract the local phase $\phi(x)$ (see Fig.\ \ref{fig:fringe}), which reveals the specific realization of the fluctuations of the \textit{in situ} local relative phase between the pair of 2D gases.
The two-point phase correlation function is defined as
\begin{equation} \label{eq:corrmap}
C_{\mathrm{exp}}(x,x') = \frac{1}{N_r} \sum_j e^{i [\phi(x)-\phi(x')]},
\end{equation}
where the index $j$ runs over $N_r$ individual experimental realizations.
The real part of the correlation function $C^r(x,x')=\text{Re}\left[C_{\mathrm{exp}}(x,x')\right]$, which is equal to $1$ for perfectly correlated pairs of points and $0$ for uncorrelated pairs of points, encodes the second-order correlation function 
\begin{equation}
    C(\bm{r},\bm{r}') := \frac{\langle \Psi_1(\bm{r})\Psi_2^{\dagger}(\bm{r})\Psi_1^{\dagger}(\bm{r}')\Psi_2(\bm{r}')\rangle}{\langle|\Psi_1(\bm{r})|^2\rangle\langle|\Psi_2(\bm{r}')|^2\rangle},
\end{equation}
where $\Psi_j(\bm{r})$ are the bosonic field operators at location $\bm{r}$ of clouds $j=1,2$. 
In Fig.~\ref{fig:corrfunc} A, we show an example of $C^r(x,x')$ taken with MRF-dressed 2D Bose gases \cite{Sunami2022}, showing decay at long distances.
Further quantitative analysis of the decay of correlations is possible by calculating $C(\overline{x})$ by averaging $C^r(x,x')$ over points with the same spatial separation $\overline{x} = x-x'$.
The extracted correlation function allows the identification of BKT critical point by sudden change in functional forms from power-law to exponential \cite{Berezinskii1972, Kosterlitz1973}.
Experimental measurements of $C(\overline{x})$, at temperatures below and above the BKT transition, are shown in Fig.~\ref{fig:corrfunc}B.
The data was taken for a near-homogeneous system of bilayer 2D Bose gases realized by adding a ring-shaped far-detuned dipole trap to the MRF-dressed trap \cite{Sunami2024}, showing distinct functional forms with algebraic and exponential decay.

\begin{figure}[t]
	\includegraphics[width=1.0\textwidth]{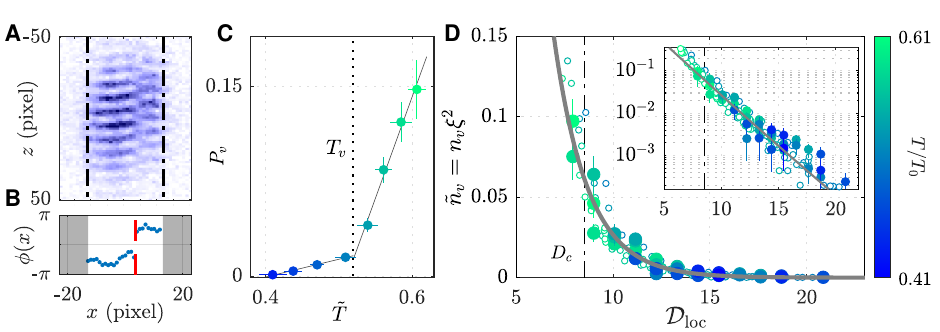}
	\caption{\label{fig:vortex} Vortex detection and local analysis of vortex density, data from Ref.\ \cite{Sunami2022}.
		(\textbf{A}) Example interference patterns with a phase dislocation.
		(\textbf{B}) The local phase extracted from fitting showing a sharp discontinuity of magnitude $\pi$, which is counted as a vortex.
		(\textbf{C}) Probability of detecting a vortex in an image $P_v$ across the BKT crossover.
		The vertical dotted line is the vortex proliferation temperature $T_v$, which is determined using a piecewise linear fit (continuous lines).
		(\textbf{D}) Scale invariant local vortex density.
		The local vortex density $\tilde{n}_v(x)=n_v(x)\xi(x)^2$ is plotted against the local PSD $\mathcal{D}_{loc}=n(x)\lambda(T)^2$, 
		where $n_v(x)$ is the local vortex density, $n(x)$ is the local 2D atom density and $\xi(x)$ is the local healing length at location $x$ in the image. 
		The measurements (filled circles) and the simulations (open circles) cover a range of temperatures across the BKT crossover and experimental datasets with eight different temperatures contribute to the plot. 
		Solid line is the exponential fit to the experimental data, and the vertical dash-dotted line is the predicted critical PSD \cite{Prokofev2001}.
		The inset shows the same results on a log-linear scale to highlight the exponential scaling across the BKT transition. Error bars are statistical.
	}
\end{figure}

\subsection{Vortex detection}

The measurement of local phase fluctuations has applications beyond averaged correlation functions.
Vortex excitations in 2D Bose gases are characterised by a winding of the phase field by integer multiples of $\pm 2 \pi$, and are often termed topological defects because of the discrete nature of phase windings which are robust against smooth deformation of the phase field.
These excitations play a crucial role in 2D systems, as the vortex-antivortex binding-unbinding is the fundamental mechanism of the BKT transition \cite{Nelson1977} that leads to the emergence of superfluidity in 2D, despite the Mermin-Wagner theorem \cite{Mermin1966, Hohenberg1967} precluding true long-range order in low-dimensional systems.
Direct detection and analysis of vortex excitations in 2D systems thus provides an insight into the microscopic mechanism of the BKT transition, which is performed by the additional analysis of matter-wave interferometry results. 
If a free, singly-charged vortex is located within the thin region where we perform the imaging, the observed phase profile displays an $\approx \pi$ phase jump as shown in Figs.~\ref{fig:vortex} A and B, where the sharp phase dislocation is indicated by a red vertical line, which we count as vortices. 
In Fig.~\ref{fig:vortex}C, we show the probability of detecting a vortex, $P_v$, measured with harmonically trapped 2D Bose gases at various temperatures $\tilde{T} = T/T_0$ ranging from 0.4 to 0.6, where $T_0=\sqrt{6N}(\hbar \omega_r/\pi k_B)$ is the condensation temperature of an ideal 2D Bose gas in a harmonic trap and the BKT critical point is measured to be at\cite{Sunami2022} $\tilde{T} = 0.53 (1)$.
There is a sharp increase in $P_v$ at a temperature indicated by $T_v$, which coincides with the BKT crossover of the system identified by the change of the functional form of the correlation function \cite{Sunami2022}.

Furthermore, the spatially resolved detection of vortices, enabled by the selective imaging, allows local analysis of vortex statistics such as shown in Fig.~\ref{fig:vortex}D.
Here, we use harmonically trapped samples where the slow variation of local density probes a range of local phase-space densities $\mathcal{D}_{loc}=n(x)\lambda(T)^2$, where $\lambda=h/\sqrt{2\pi m k_B T}$ is the thermal de Broglie wavelength.
The data suggest that the local vortex density can be inferred by the local phase space density (PSD) only, with a potential application of local density description of free vortices\cite{Hung2011}.

\begin{figure}[t]
\includegraphics[width=0.97\textwidth]{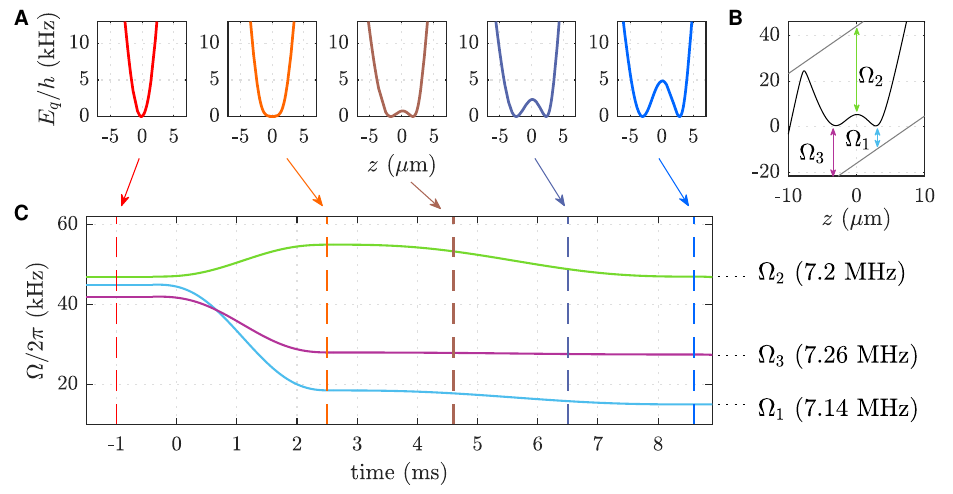}
\caption{\label{fig:splitting} Coherent splitting of two-dimensional gases in an MRF-dressed potential.
(\textbf{A},\textbf{C}) The RF amplitude ramp, expressed in terms of Rabi frequencies for each RF component $\Omega_i$, is plotted together with the potential shape at selected times (dashed vertical lines).
Panel (\textbf{B}) illustrates the effect of the RF amplitudes to trap geometry plotted including the effect of gravity in the $-z$ direction. Straight lines are the eigenenergies of $\widetilde{m}_F = 0$ states at different RF photon Fock states.
As indicated, the gravitational sag is compensated throughout the entire process to give coherent splitting into two balanced wells by the precise calibration of the RF amplitude ramps. 
The adiabatic loading into MRF-dressed double well, described in Sec.~\ref{sec:impl}, is performed with a similar temporal profile of the RF amplitudes, however with a timescale that is up to 100 times longer.
}
\end{figure}

\section{\label{sec:quench} Coherent splitting quench of 2D quantum gas}
MRF-dressed traps enable dynamical control over the shape of the double-well potential by changing the amplitude and phase of the dressing RF field. 
An example of such protocol is the coherent splitting of a single 2D gas into two 2D systems, which causes a rapid change of atom density.
This is similar to the method for splitting 1D quantum gases with an atom chip to study dynamics of near-integrable systems \cite{Gring2012, Langen2015, Rauer2018, Schweigler2017}. However, as two-dimensional systems possess the BKT critical point, this enables the investigation of universal scaling dynamics \cite{Sunami2023, Mathey2017} of the BKT universality class.
For concreteness, below we describe the MRF-dressing based quantum gas splitting procedure for 2D Bose gases, while this experimental technique is applicable for other geometries with different static field configurations.

The splitting procedure starts with a 2D quantum gas trapped in a three-RF single-well trap as shown in Fig.~\ref{fig:splitting}A (red, left) with the corresponding RF amplitudes shown in Fig.~\ref{fig:splitting}C at the red vertical dashed line. 
Subsequently, $\Omega_2$ is increased while $\Omega_{1}$ and $\Omega_{3}$ are reduced, to realize a flat-bottom potential with vanishing quadratic term shown in Fig.~\ref{fig:splitting}A (orange) in preparation for the splitting procedure.
\begin{figure}[t]
\includegraphics[width=0.88\textwidth]{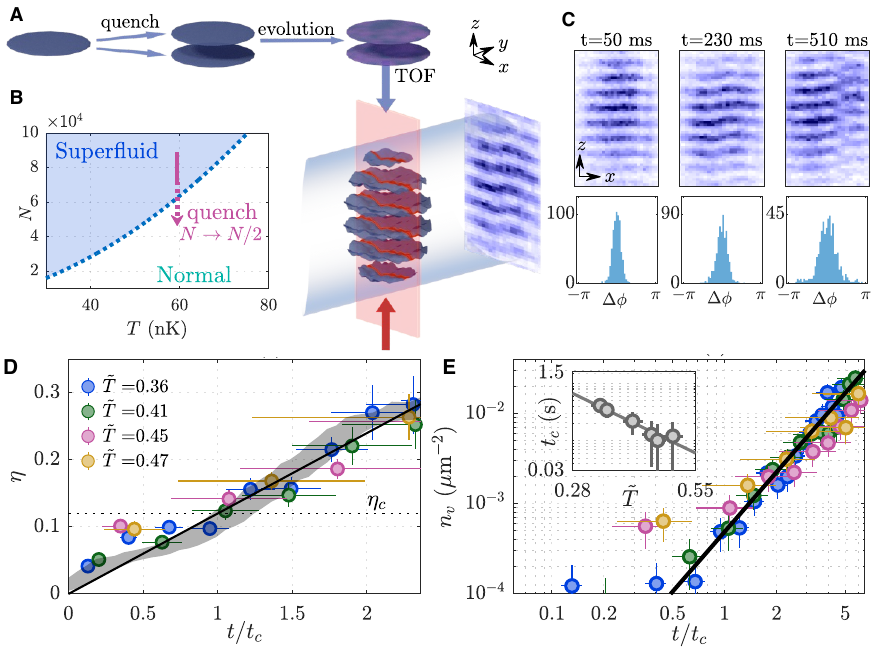}
\caption{\label{fig:quenchFig} Observation of non-equilibrium dynamics in 2D Bose gases via matter-wave interferometry, data taken from Ref.\ \cite{Sunami2023}.
    (\textbf{A}) A 2D superfluid is split into two daughter clouds, thereby quenching through the BKT transition.
    The two clouds evolve for time $t$ and are released to produce matter-wave interference after a time-of-flight (TOF).
    Local phase fluctuations are observed by optically pumping the slice (red sheet) and then performing absorption imaging.
    (\textbf{B}) Equilibrium phase-diagram of trapped 2D Bose gases \cite{Sunami2022, Holzmann2008}. 
    The quench forces the system out of equilibrium towards the normal phase. 
    (\textbf{C}) Examples of interference images (top).
    Phase dislocation caused by a vortex is visible in the image at 510 ms.
    The histograms (bottom) show the phase differences $\Delta \phi = \phi(x)-\phi(x')$ at $|x-x'|=5\,\mu$m from 45 experimental runs.
    The decreasing height and increasing width indicate increased phase fluctuations. 
    (\textbf{D}) Time evolution $\eta(t/t_c)$ scaled according to the $\tilde{T}$-dependent crossover time $t_c$.
    The horizontal error bars arise from the uncertainty in $t_c$.
    Critical exponent $\eta_c =0.13(1)$ (horizontal dotted line) is obtained at $t/t_c=1$.
    The gray shaded curve indicates the simulation result at and the solid line is a guide to the eye.
    (\textbf{E}) Scaled time evolution $n_v(t/t_c)$, plotted on a log-log scale, displays a universal growth after $t_c$.
    Black solid line is the fit with power-law $n_v \propto t^{2\nu}$ which yields $\nu=1.1(1)$.
    The inset shows the dependence of $t_c$ on $\tilde{T}$ with a solid line as a guide to the eye.
}
\end{figure}
The balanced splitting into a double well is performed by reducing $\Omega_2$ while changing $\Omega_{1}$ and $\Omega_{3}$ slowly to keep the wells balanced. The final double-well has a well separation of $\gtrsim \SI{6}{\micro \meter}$ and a barrier height of $\gtrsim 4$ kHz, sufficient to decouple\footnote{We perform further decoupling sequence to ensure complete decoupling of the two layers for experiments involving long probe time of $\sim 1$ s} the two clouds which have a thermal energy scale $k_B T$ and a chemical potential $\mu$ on the order of $\lesssim 1$ kHz.
%changed to 0.1 um. At 1 kHz, harmonic oscillator length is only a few tens of nm. 
As illustrated in Figs.~\ref{fig:quenchFig}A and B, the splitting procedure realizes a \textit{quench} from the superfluid to the normal phase to prepare repeatable out-of-equilibrium states from which the relaxation dynamics can be monitored by matter-wave interferometry\cite{Mathey2017,Barker2020,Murad2022,Sunami2023}.
Following the quench, the system relaxes towards the vortex-dominated normal phase, as indicated by the increase of power-law exponent $\eta$ characterizing the decay of correlation function and the vortex density $n_v$ (Figs.~\ref{fig:quenchFig}D and E). 
Scaled by the characteristic timescale for each dataset at different initial temperature $t_c(\tilde{T})$ (Fig.~\ref{fig:quenchFig}E inset), the time at which the correlation function becomes better described by an exponential model, the dynamics at different initial conditions collapse onto a unique time trajectory \cite{Sunami2023}.
In particular, the power-law increase of vortex density $n_v(t) \propto t^{2}$ (Fig.~\ref{fig:quenchFig}E, solid black line) indicates a scaling behavior which is consistent with the real-time renormalization-group (RG) theory of the relaxation dynamics \cite{Mathey2010,Mathey2017}.
Further comparison of the experimental data with the RG theory was performed \cite{Sunami2023}, supporting the viability of RG-based interpretation for understanding the universal critical dynamics.

\begin{figure}[t]
\includegraphics[width=0.99\textwidth]{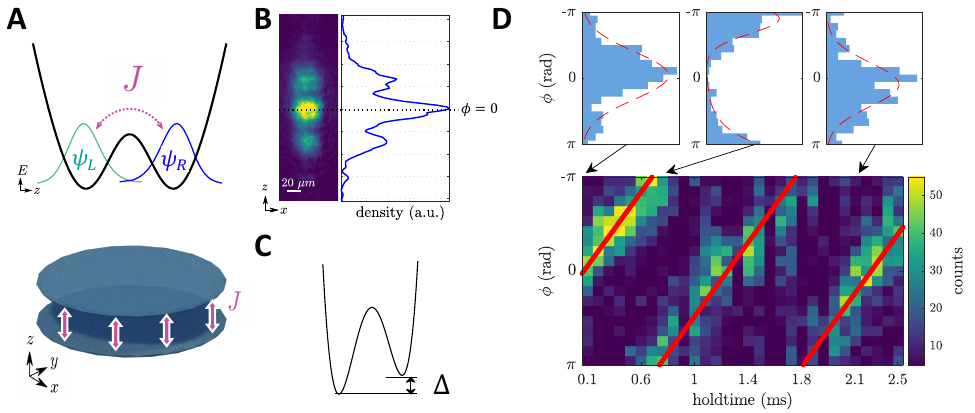}
\caption{\label{fig:coupled} Realization and control of coupled bilayer 2D systems.
(\textbf{A}) Illustration of double-well trap with tunnel-coupling $J$. For a bilayer system, the coupling occurs over the entire surface as indicated by arrows. 
Control over $J$ can be attained by the appropriate selection of RF frequencies that create the MRF-dressed potential.
(\textbf{B}) The interference fringes of the bilayer system with small well separation of $\lesssim \SI{2}{\micro\meter}$. 
The interference fringes have a longer wavelength compared to the experiments with large well separations, such as the data shown in Fig.~\ref{fig:fringe}.
(\textbf{C}) Decoupled, imbalanced bilayer system is the starting point for the observation of the phase oscillation in coupling-quenched systems \cite{Luick2020, Pigneur2018}. 
(\textbf{D}) Phase accumulation sequence in imbalanced, decoupled systems \cite{SunamiThesis}, demonstrating the stability of the apparatus to probe the relative phase dynamics.
The system is initialized by a coherent splitting without crossing the critical point, resulting in two superfluids with relative phase $ \phi = 0$.
The histogram of the extracted phase $\phi$ of the interference fringes, at each hold time, are plotted as a vertical strip (corresponding histograms are shown as top panels for three selected times, where red dashed lines are the Gaussian fit), where colors indicate the occurrence.
Red solid lines are the relative phase accumulation predicted by the well imbalance $\Delta \sim \SI{1}{\kilo \hertz}$ of the double-well trap.
}
\end{figure}

\section{\label{sec:outlook} outlook}
In this perspective, we provided an overview of the MRF dressing techniques and their application to the investigation of BKT physics by controllable bilayer 2D quantum gases. 
The MRF-dressed potential approach extends the rich technical capability of RF-dressed traps by making use of the spectral degrees of freedom for flexible spatial trap engineering.
In addition to the further studies of coupled bilayer 2D Bose gases, the MRF technique has much wider applications including other RF-dressed trap topologies and species-selective control and detection, which we outline in this concluding section.

\subsection{\label{sec:coupled_bilayer}Inter-layer coupling control and bilayer Josephson dynamics}
The system of two superfluids with weak tunnel-coupling has similarities with a superconducting Josephson junction \cite{Zapata1998,Davis2002}, and the atomic system realizations are called bosonic Josephson junction (BJJ) and have been studied extensively for 1D and 3D systems \cite{Gross2010,Albiez2005,Pigneur2018}, as well as single-layer 2D systems split in half by a potential barrier \cite{Luick2020}.
Bilayer 2D double-well systems present a novel structure of the BJJ, with unique opportunity to probe the physics of layered superfluids, such as counterflow superfluidity \cite{Homann2024}, light-induced superconductivity in a driven junction \cite{Okamoto2016, Zhu2021}, as well as the novel class of phase transitions in strongly coupled 2D systems \cite{Mathey2007, Song2021}.
The MRF-dressed potential allows precise and highly stable production of layered Josephson junctions, as the coupling can be controlled by the RF frequency separations, extremely stable thanks to modern RF sources, that control the spatial separation of the wells.
The Josephson coupling strength $J$ has approximately exponential dependence on the well separation, from the Gaussian ground-state wavefunction in each well \cite{Anakian2006}.
Loading of ultracold atoms to a bilayer trap with well separation of $\lesssim \SI{2}{\micro\meter}$ has been demonstrated \cite{SunamiThesis}, with which coupling strength $J/h$ on the order of $10 - 100 $ Hz are expected.
This is a promising step towards the observation of strong phase-locking phenomena and its effect on the bulk properties of 2D quantum systems \cite{Mathey2007, Bighin2019, Song2021, Eto2018, Kobayashi2019}.

\subsection{MRF dressing for other geometries}

\begin{figure}[t]
\includegraphics[width=1.0\textwidth]{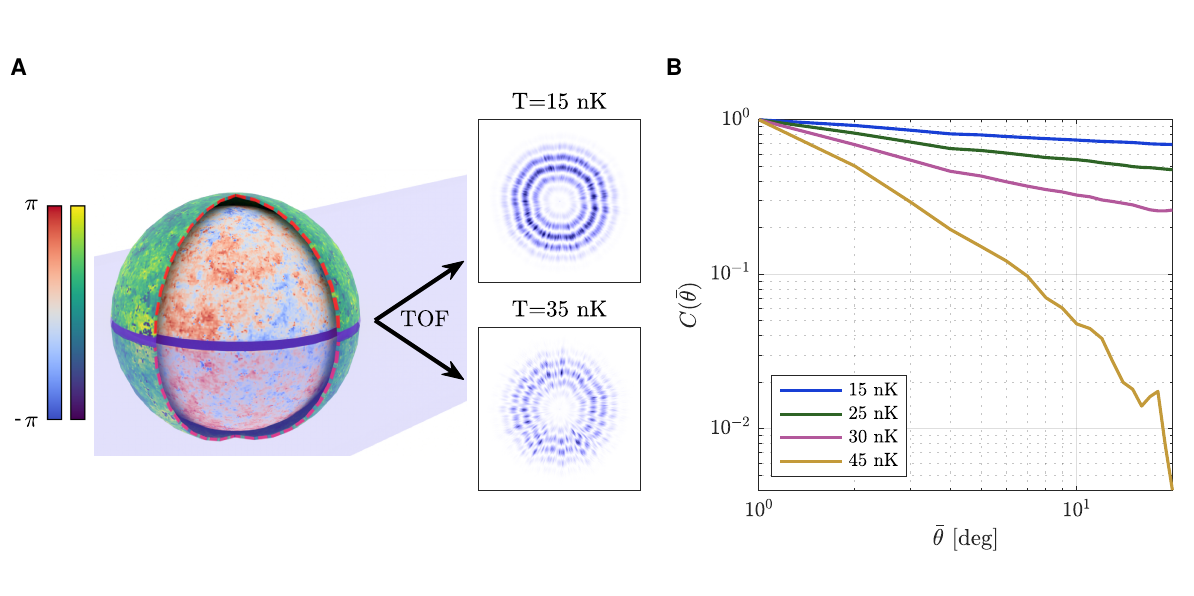}
\caption{
(\textbf{A}) Illustration of a possible experiment for the matter-wave interferometry of RF-dressed bilayer bubble trap.
The phase fluctuations are imprinted onto the bubble-shaped clouds and illustrated by two different color maps for the two layers (the outer layer is partially cut away for illustration purposes). 
Purple shading indicates the region of selective imaging, which is performed in the vertical direction.
Simulated matter-wave interference patterns, using numerical Gross-Pitaevskii simulation of expansion dynamics, displays larger phase fluctuations at higher temperatures.
(\textbf{B}) Correlation functions extracted from simulated time-of-flight images with $C(\bar{\theta}) = \langle e^{i(\phi(\theta)-\phi(\theta+\bar{\theta})} \rangle$, where $\phi(\theta)$ is the local phase of the radial interference pattern at polar angle $\theta$. The correlation function changes from algebraic to exponential as the system crosses the superfluid transition. 
}
\label{fig:bubble_fringes} 
\end{figure}

RF-dressed trapping has already been realized in various interesting topologies, including Ioffe field \cite{Colombe2004}, atom chip \cite{Schumm2005}, ring trap \cite{Morizot2006,Heathcote2008, Navez2016}, optical lattice \cite{Lundblad2008}, and bubble traps in space-based experiments \cite{Carollo2022}. 
Employing multiple radio frequencies extends the controllability of the trapping potential, allowing both much in-depth study of the many-body systems and advancement of the technical capability of atom interferometers.
In particular, the superfluid transition is an interesting topic on curved and closed manifolds \cite{PhysRevResearch.4.013122}. 
While detection of this phase transition is expected to be possible by measuring the sound modes, direct measurement of phase fluctuations offers alternative probe with in-depth understanding of correlation properties and underlying mechanism of vortex excitations.
In Fig.~\ref{fig:bubble_fringes}, we illustrate a potential scheme to probe the phase fluctuations of bubble-shaped quasi-2D systems, based on the MRF-dressing technique.
The ultracold atoms are loaded into the single-RF shell trap, which is then adiabatically loaded into a double-well potential, which interfere after the sudden turn-off of the trap.
Spatially selective imaging allows the direct detection of radially encoded interference pattern which can be analyzed (see Fig.~\ref{fig:fringe}) to obtain the phase fluctuations.
Performing phase extraction and the associated phase correlation analysis in a similar manner to the planar case (see Fig.~\ref{fig:corrfunc}), would result in the correlation functions as shown in Fig.~\ref{fig:bubble_fringes}B, showing the power-law to exponential crossover as the system approaches high-temperature phase.

\subsection{MRF-dressed mixture: species-selective manipulation and detection}
The MRF-dressing technique allows individual trapping and manipulation of mixtures of quantum gases with differing Land\'e $g$-factors \cite{Bentine2017, Bentine2020, Barker2020jphysb}.
For example, the two hyperfine ground levels of Rubidium have $g$-factors of opposite sign, allowing individual control by RF polarization \cite{Barker2020jphysb}.
For MRF-dressed potentials, phase engineering allows trapping of each species in largely different trap geometry \cite{Barker2020jphysb}, or to perform selective trap switching for selective time-of-flight imaging.
This allows the application of interferometry-based probing of many-body systems to multi-component systems, which provides an opportunity to further engineer the Hamiltonian with a wide variety of parameters such as tunable interaction mediated by the second component, coupled four-component system in a double-well \cite{Barker2020jphysb}, as well as a design of more flexible interferometers \cite{Barker2020jphysb, Jammi2018a, IloOkeke2021}, while utilizing the powerful spatial interferometry probe of 2D systems.

\section*{journal credit line}
The following article has been submitted to AVS Quantum Science. After it is published, it will be found at https://pubs.aip.org/avs/aqs.

\section*{Acknowledgements}
We greatly acknowledge our theory collaborators, especially V.~Singh and L.~Mathey, for their contribution to the inception and interpretation of quench experiments, and thank all former and present members of Oxford Quantum Matter Laboratory who contributed to the development of the MRF-dressing experiment.
This work was supported by the EPSRC Grant Reference EP/X024601/1. 

\section*{Author Declarations}
\subsection*{Conflict of Interest}
The authors have no conflicts to disclose.

\subsection*{Data availability}
The data that support this study are available from the corresponding author upon reasonable request.

\end{document}